\documentclass[preprint,nofootinbib,aps,superscriptaddress,eqsecnum]{revtex4-1} 
\pdfoutput=1
\usepackage{rotating}
 \usepackage{graphicx}
 \usepackage{xcolor}
 \usepackage{amsmath}
\usepackage{amsfonts}
\usepackage{subfigure}
\usepackage{amssymb}
\usepackage{hyperref}
\usepackage{gensymb}
\usepackage{slashed}
\usepackage{caption}
\usepackage{ulem}
\def\bea{\begin{eqnarray}}
\def\eea{\end{eqnarray}}
\def\be{\begin{equation}}
\def\ee{\end{equation}}

\def\f{\mathcal{E}}
\def\g{\mathcal{B}}

\def\D{\text{d}}
\begin{document} 
\title{Shadows in dyonic Kerr-Sen black holes}
\author{Soumya Jana }
\email[Email Address: ]{soumyajana.physics@gmail.com}
\affiliation{Department of Physics, Sitananda College, Nandigram, 721631, India}
\affiliation{Department of Physics, Indian Institute of Technology, Kharagpur - 721302, India}
\author{Sayan Kar}
\email[Email Address: ]{sayan@phy.iitkgp.ac.in }
\affiliation{Department of Physics, Indian Institute of Technology, Kharagpur - 721302, India}
\begin{abstract}
Black holes with dyonic charges in Einstein-Maxwell-dilaton-axion supergravity theory are revisited in the context of black hole shadows. We consider static as well as rotating (namely the dyonic Kerr-Sen) black holes. The matter stress-energy tensor components, sourced by the Maxwell, axion and dilaton fields satisfy the standard energy conditions. The analytical expressions for the 
horizon and the shadow radius of the static 
spacetimes demonstrate their dependence on $P^2+Q^2$ ($P$, $Q$ the magnetic and electric charges, respectively) and the mass parameter $M$. 
The shadow radius lies in the range $2M <R_{shadow}<3\sqrt{3} M$
and there is no stable photon orbit outside the horizon. Further, shadows cast by the rotating dyonic Kerr-Sen black holes are also studied and compared graphically with 
their Kerr-Newman and Kerr-Sen counterparts. Deviation of the shadow boundary is prominent with the variation of the magnetic charge, for the relatively slowly rotating dyonic Kerr-Sen spacetimes. We test any possible presence of a magnetic monopole charge in the backdrop of recent EHT observations for 
the supermassive black holes M87$^*$ and Sgr A$^*$. Deviation from circularity of the shadow boundary ($\Delta C$) and deviation of the average shadow radius from the Schwarzschild shadow radius (quantified as the fractional deviation parameter $\delta$) are the two observables used here. The observational bound on $\Delta C$ (available only for M87$^*$) is satisfied for all theoretically allowed regions of parameter space and thus cannot constrain the parameters. The observational bound on $\delta$ available for Sgr A$^*$ translates into an upper limit on 
any possible magnetic monopole charge 
linked to Sgr A$^*$ and is given as $P\lesssim 0.873 M$. 
Constraints on $P$ obtained  from other astrophysical
effects are however expected to be far 
more stringent though rigorous analyses 
along these lines is lacking in the literature. 
In addition, future refined imaging (shadow) observations 
will surely help in improving the bound on $P$ arrived at here.

\end{abstract}

\maketitle
\section{Introduction}
The Reissner-Nordstr\"om (RN) geometry representing the gravitational field due
to a charged massive object is among the earliest known exact solutions
in general relativity coupled to electromagnetism, i.e. the Einstein-Maxwell theory.
A straightforward generalization of this solution is the dyonic RN spacetime
\cite{dyonicRN} which can be written down by just replacing the $Q^2$ in RN
spacetime with $P^2+Q^2$, where $P$ represents the ``magnetic" charge and
$Q$ is its ``electric" counterpart. However, for the dyonic solution, the
definition of the electromagnetic potential $A_i$ is a little tricky
-one needs, as expected for a dyon, a two-patch definition - one for the
northern hemisphere and the other for the southern. The standard electric-magnetic
duality which arises when both magnetic and electric charges are present keeps
the solution unchanged. The horizons and other features for the dyonic RN spacetime
resemble those for the usual RN geometry modulo the presence of the extra
magnetic charge.

It is also known that additional matter fields (other than Maxwell) such as the dilaton and/or the
axion appear in the context of supergravity theories or in low energy effective
actions which emerge out of full string theory \cite{Gibbons:1987ps,Garfinkle:1990qj,Sen:1992ua}. 
In such scenarios too one expects dyonic solutions representing gravitational
fields of such objects. Among various known solutions \cite{Gibbons:1987ps,Garfinkle:1990qj,Campbell:1991rz,Sen:1992ua,Shapere:1991ta,PhysRevD.48.742} there are static, spherically
symmetric ones as well as stationary spacetimes wherein rotation is present.
The purpose of this article is to revisit such known solutions without and with
rotation. Our primary aim is to learn how the various theory parameters (e.g. electric, magnetic
as well as other charges) control the nature and profile of the shadow/silhouette
created by the gravitational field representing such solutions. We also try and
see if any meaningful constraint can be placed on the various charges, using 
the known shadow observations for the supermassive compact object present in
M87$^*$ \cite{Akiyama_2019L1,Akiyama_2019L5,Akiyama_2019L6} and for Sgr A* \cite{EHT2022_1,EHT2022_6}. Though, dyonic scenarios presently have little to do with observations
in other contexts, we will see how one may place bounds on their viability 
through shadow observations. In other words, we try to show that what is seen
in the images may also be explained using hypothetical constructs which, 
by no means can be ruled out altogether, unless other observations
imply mismatches and contradictions.

Shadows in Kerr-Sen black holes \cite{Sen:1992ua} have already been studied by several authors \cite{Hioki:2008zw,Xavier:2020egv,Narang:2020bgo}. The rotating version of the dyonic black holes in Einstein-Maxwell-dilaton theory \cite{Gibbons:1987ps,Cheng:1993wp} and its shadows was studied in \cite{Shaikh:2019fpu}. In \cite{Banerjee:2022bxg}, the authors investigated shadows of regular (Bardeen) black holes having magnetic monopole charge sourced by nonlinear electrodynamics coupled to GR. The presence of axionic hair or the Kalb-Ramond field and their influence on the shadow of M87$^*$ was investigated in \cite{Banerjee:2019xds}. In \cite{PhysRevD.103.044057}, the authors investigated the effect of QED on the shadows of the static black holes with magnetic monopoles.  There are also several other studies on the shadows of hairy black holes. For example, the authors of \cite{PhysRevD.103.064026} studied the shadows cast by the rotating black holes with anisotropic matter fields which could describe an extra $U(1)$ field as well as diverse dark matter. Studies on the shadows of braneworld blackholes such as in \cite{PhysRevD.101.041301, PhysRevD.104.024001} are among other examples. For a more recent study on the shadows of the black holes in the extended or alternative gravity theories, in the light of the observations of Sgr A$^*$, see \cite{Vagnozzi:2022moj}.

Our article is organized as follows. In Sec. II we recall the static black hole  solutions and discuss the energy conditions in Einstein-Maxwell--dilaton-axion (EMDA) supergravity theory. Section III provides a summary of the corresponding stationary solutions 
(dyonic) which
include rotation. Shadow calculations and related details are presented in Sec. IV and connections/comparisons with observations are outlined in Sec. V. Sec. VI is a summary with concluding remarks. 

Throughout the paper we consider the units $G=1$ and $c=1$.

\section{Black holes in Einstein-Maxwell--Dilaton-Axion supergravity theory}
The Einstein-Maxwell -dilaton- axion (EMDA) supergravity theory is described by the action \cite{Shapere:1991ta}
\begin{equation}
    S_{\text{EMDA}}= \int \D^4 x \sqrt{-g}\left[ R - \frac{1}{2}(\partial \Phi)^2-\frac{1}{2}e^{2\Phi}(\partial \xi )^2 -e^{-\Phi}F^2 + \xi F_{\mu}\tilde{F}^{\mu\nu}\right]
    \label{EMDA action}
\end{equation}
where $\Phi$ and $\xi$ are the dilaton and axion fields, $F_{\mu\nu}$ is the usual electromagnetic field tensor, $F^2= F_{\mu\nu}{F^{\mu\nu}}$, and $\tilde{F}^{\mu\nu}=\frac{1}{2\sqrt{-g}}\epsilon^{\mu\nu\alpha\beta}F_{\alpha\beta}$ is the dual electromagnetic field tensor. The equations of motion of the dilaton, axion fields are obtained as
\begin{equation}
    \Box \Phi -e^{2\Phi}(\partial \xi)^2 + e^{-\Phi}F^2 = 0,
    \label{eom dilaton}
\end{equation}
and
\begin{equation}
    \Box \xi + 2\nabla^{\mu}\Phi\nabla_{\mu}\xi + e^{-2\Phi}F_{\mu\nu}\tilde{F}^{\mu\nu} = 0.
    \label{eom axion}
\end{equation}
The equation of motion for the electromagnetic vector potential $A_{\mu}$ ( where $F_{\mu\nu}= \partial_{\mu}A_{\nu}- \partial_{\nu}A_{\mu}$) is obtained as
\begin{equation}
    \nabla_{\mu}\left(-e^{-\Phi}F^{\mu\nu}+ \xi \tilde{F}^{\mu\nu}\right)= 0,
    \label{eom A}
\end{equation}
along with the usual Bianchi identity,
\begin{equation}
    \nabla_{\mu} \tilde{F}^{\mu\nu}=0.
    \label{bianchi}
\end{equation}
The equation of motion for the metric tensor $g_{\mu\nu}$ is obtained by varying the action (\ref{EMDA action}) with respect to $g_{\mu\nu}$. We get
\begin{equation}
    R_{\mu\nu}= \frac{1}{2}\nabla_{\mu}\nabla_{\nu}\Phi + \frac{1}{2}e^{2\Phi} \nabla_{\mu} \xi \nabla_{\nu} \xi + 2 e^{-\Phi}F_{\mu\alpha}F_{\nu}{}^{\alpha} - \frac{1}{2}g_{\mu\nu}e^{-\Phi}F^2,
    \label{ricci eqs}
\end{equation}
where $R_{\mu\nu}$ are the Ricci tensor components.

\subsection{Static black hole solution}
The static black hole solution in such a system has already been obtained in  \cite{Shapere:1991ta}, where the authors used symmetry transformations on the axion and dilaton fields to obtain its form. In this section, we first outline the derivation of the same solution by solving directly, the Einstein field equations. Thereafter,  we analyze the structure of the black hole solution.

We consider the ansatz for the spherically symmetric static line element
\begin{equation}
\D s^2_{static}= -\Delta^2(R)\D t^2 + \frac{\psi^2(R)}{\Delta^2(R)} \D R^2 + R^2\left( \D \theta^2 + \sin^2 \theta \D \phi^2\right).
\label{eq:static line element}
\end{equation}
We also assume nonvanishing components of the electromagnetic field tensor 
\begin{equation}
    F_{01}=- F_{10}= \f(R), \, \quad~ F_{23}= -F_{32} = \g(R)\sin \theta.
\end{equation}
Then Eq.~(\ref{eom A}) becomes
\begin{equation}
    \left(\frac{e^{-\Phi}\f R^2}{\psi}+ \xi \g \right)^{\prime}=0,
    \label{eom A sph}
\end{equation}
where the prime (${}^{\prime}$) denotes the derivative with respect to the radial coordinate $R$. The Bianchi identity is satisfied for $\g(R)=P$ (a constant) and  $P$ is therefore identified as the magnetic charge since
\begin{equation}
    \tilde{F}^{10}= -\frac{P}{R^2\psi}.
\end{equation}
Integrating Eq.~(\ref{eom A sph}), we get
\begin{equation}
    \f= \psi e^{\Phi}\left(\frac{Q- \xi P}{R^2}\right),
\end{equation}
where the integration constant $Q$ is identified as the electric charge, since, at large $R$, $\f\sim Q/R^2$.
The equations of motion [Eqs.(\ref{eom dilaton})-(\ref{eom axion})] for the dilaton and the axion field become
\begin{equation}
    \left(\frac{\Delta^2R^2\Phi^{\prime}}{\psi}\right)^{\prime}= 2 e^{-\Phi}\psi R^2 \left(\frac{\f^2}{\psi^2}-\frac{P^2}{R^4}\right) + e^{2\Phi} \frac{R^2\Delta^2}{\psi}\xi^{\prime}{}^2,
    \label{eom dilaton sph}
\end{equation}
and
\begin{equation}
    \left(\frac{\Delta^2R^2\xi^{\prime}}{\psi}\right)^{\prime} = - 2\frac{\Delta^2R^2}{\psi}\Phi^{\prime}\xi^{\prime} - 4e^{-2\Phi}P\f. \label{eom axion sph}
\end{equation}
Note from Eq.~(\ref{eom axion sph}) that for $\xi =0 $, $P=0$ or $Q=0$.
From Eq.~(\ref{ricci eqs}), we get three nonvanishing components which are
\begin{eqnarray}
    R_{00}&=& \frac{\Delta^4}{\psi^2}\left[\frac{\Delta^{\prime}{}^2}{\Delta^2} - \frac{\Delta^{\prime}\psi^{\prime}}{\Delta \psi } + \frac{(\Delta^{\prime}R^2)^{\prime}}{\Delta R^2}\right]= e^{-\Phi}\Delta^2 \left(\frac{\f^2}{\psi^2}+\frac{P^2}{R^4}\right), \label{eq:R00}\\
    R_{11}&=& \frac{2\psi^{\prime}}{R\psi} -\frac{\Delta^{\prime}{}^2}{\Delta^2} + \frac{\Delta^{\prime}\psi^{\prime}}{\Delta \psi } - \frac{(\Delta^{\prime}R^2)^{\prime}}{\Delta R^2}= \frac{1}{2}\Phi^{\prime}{}^2 + \frac{1}{2}e^{2\Phi}\xi^{\prime}{}^2- e^{-\Phi}\frac{\psi^2}{\Delta^2} \left(\frac{\f^2}{\psi^2}+\frac{P^2}{R^4}\right), \label{eq:R11}\\
    R_{22}&=& 1- 2R\frac{\Delta \Delta^{\prime}}{\psi^2} + \frac{\Delta^2R^2}{\psi^3}\left(\frac{\psi}{R}\right)^{\prime} = R^2e^{-\Phi}\left(\frac{\f^2}{\psi^2}+\frac{P^2}{R^4}\right). \label{eq:R22}
\end{eqnarray}
Using Eqs.~(\ref{eq:R00}) and (\ref{eq:R11}), we get 
\begin{equation}
    \frac{2}{R}\frac{\psi^{\prime}}{\psi} = \frac{1}{2}\Phi^{\prime}{}^2 + \frac{1}{2}e^{2\Phi}\xi^{\prime}{}^2.
\end{equation}
Demanding proper asymptotic behaviour, i.e. $\Phi\rightarrow 0$ and $\xi \rightarrow 0$, $\psi\rightarrow 1$ for $R\rightarrow \infty$, we assume, 
\begin{equation}
    \frac{\psi^{\prime}}{\psi}= \frac{\sigma^2}{R\left(R^2+\sigma^2\right)},\quad~ \text{or,} \quad~ \quad~ \psi^2= \frac{R^2}{R^2+\sigma^2},
    \label{eq:psi_anstaz}
\end{equation}
where $\sigma^2$ is a constant. From Eqs.~(\ref{eq:R00}) and (\ref{eq:R22}), we get
\begin{equation}
    \left[\frac{1}{2}\frac{\left(\Delta^2 R^2\right)^{\prime}}{\psi}\right]^{\prime}= \psi.
\end{equation}
Solving this equation with the assumption on $\psi$ given in Eq.~(\ref{eq:psi_anstaz}), we obtain the solution for $\Delta(r)$ as
\begin{equation}
    \Delta^2(R)= 1- \frac{2M\sqrt{R^2+\sigma^2}}{R^2} + \frac{P^2+Q^2}{R^2},
    \label{eq: Delta}
\end{equation}
where the integration constants are identified as the mass $M$ and the sum of the square of the charges ($P^2+Q^2$). The black hole resembles the Reissner-Nordstr\"om black holes asymptotically. Using Eqs.~(\ref{eq:psi_anstaz}) and (\ref{eq: Delta}) the solution for the equations of motion for the dilaton and axion fields is
\begin{eqnarray}
e^{\Phi}&=& 1 + \frac{2d}{R^2}\sqrt{R^2+k^2+d^2} + \frac{2(k^2+d^2)}{R^2},\\  
\xi &=& \frac{2 k \sqrt{R^2+k^2 + d^2}}{R^2 +2 d \sqrt{R^2+ k^2 + d^2} + 2(k^2+ d^2)},
\end{eqnarray}
 where, $d= \frac{(P^2-Q^2)}{2M}$ and $k= \frac{PQ}{M}$ are dilaton and axion charges, respectively and $\sigma^2= k^2 + d^2$.

\begin{figure}[!htbp]
\centering
\includegraphics[width=3.0in,angle=360]{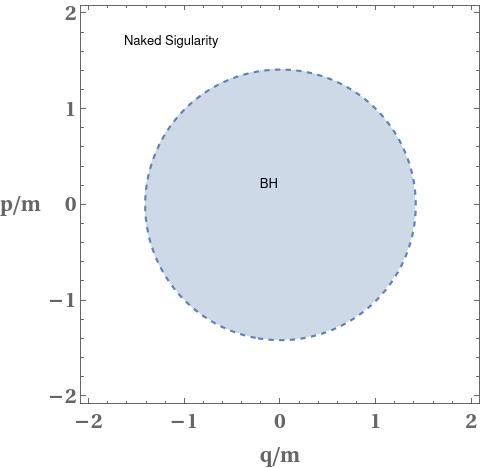}
\caption{$\frac{P}{M}$ vs $\frac{Q}{M}$ parameter space is plotted. The shaded circular region indicates the allowed parameter space for black holes and the exterior region in parameter space corresponds to naked singularities. }
\label{fig:f2}
\end{figure}

In the absence of both electric and magnetic charges (i.e. $P=Q=0$), the dilaton and axionic charges also vanish, i.e. $k=d=0$, and we recover the Schwarzschild black hole.  For any nonzero $P$ and/or $Q$, the line element does not resemble the Riessner-Nordstr\"om black holes. This signifies that these black holes are hairy. Another distinguishing feature of these black holes is that they have single horizons -- unlike the Riessner-Nordstr\"om. Using the relation $k^2+d^2=(P^2+Q^2)^2/{4M^2}$ in the $f(R_{hz})=0$, we get the horizon radius as,
\begin{equation}
    R_{hz}=2M\sqrt{1-\frac{P^2+Q^2}{2M^2}}.
\end{equation}
This feature is also different from  Riessner-Nordstr\"om and the dyonic black holes with dilaton field as the only scalar hair. We notice a double horizon in general and extremal horizon in a special situation. However, the  static version of the Kerr-Sen black holes shares the similar feature of a single horizon. For black holes,
\begin{equation}
    M\geq \sqrt{\frac{P^2+Q^2}{2}},
\end{equation}
otherwise, we have naked singularities. This is illustrated in Fig.~\ref{fig:f2}.
 
\subsection{Energy conditions}
By identifying the nonzero components of the stress-energy tensors as $T^0{}_0=- \rho$, $T^1{}_1= \tau$, and $T^2{}_2=T^3{}_3=p$, where $\rho$ (energy density), $\tau$ (radial pressure), and $p$ (tangential pressure) are defined in the orthonormal frame basis,  and using the Einstein field equations ($T_{\mu\nu}=G_{\mu\nu}/{8\pi G}$), we analyze all the energy conditions. We find that:

$(a)$ The Null Energy Conditions (NEC), i.e. $\rho +\tau \geq 0$ and $\rho + p\geq 0  $ are satisfied when $\Delta^2(R)\geq 0$. This implies that the NEC is satisfied on and outside the horizon of the black hole. For a naked singularity, the NEC is satisfied for all $R$.

$(b)$ The Weak Energy Conditions (WEC) implies $\rho\geq 0$ in addition to the NEC. Using the Einstein field equation
\begin{equation}
    \rho = \frac{1}{8\pi G}\left[\frac{1}{R^2}-\frac{(\Delta^2)'}{R\psi^2}-\frac{\Delta^2}{R^2\psi^2}+2\frac{\Delta^2\psi'}{R\psi^3}\right].
\end{equation}
One can check that $\rho\geq 0$ for $R\geq R_{hz}/\sqrt{3}$ for black holes, and for all $R$ in case of naked singularities. Thus, the WEC is also satisfied on and outside the horizon of the black holes.

$(c)$ The Strong Energy Condition (SEC), i.e. $\rho+\tau+2p \geq 0$ is satisfied for all $R$ irrespective of black holes or naked singularities.

$(d)$ The Dominant Energy Conditions (DEC), i.e. $\rho\geq 0$, $\rho\geq \vert\tau \vert$, and $\rho \geq \vert p \vert$ are satisfied on and outside of the black hole horizon, and for all $R$ in case of naked singularities.  

It was conjectured \cite{PhysRevD.94.106005}, \cite{Guo:2022ghl} that ``a violation of either the dominant or the strong energy
condition is a necessary condition for the existence of an antiphoton sphere outside a regular black
hole horizon". Thus, according to our analysis of the energy conditions, the static dyonic black holes in the EMDA theory do not consist antiphoton sphere or stable photon orbits. 

\section{Dyonic Kerr-Sen black holes}
Dyonic Kerr-Sen black holes are the rotating versions of the static black holes. The Newman-Janis (NJ) algorithm \cite{Newman:1965tw,Newmn:1965b} can be applied to obtain such rotating black holes. After introducing a new radial coordinate $r$ such that the squared area radius $R^2=r^2-2dr-k^2$ or $r=\sqrt{R^2+k^2+d^2} +d $, the static line element~(\ref{eq:static line element},\ref{eq:psi_anstaz}, \ref{eq: Delta}) becomes  
\begin{equation}
    ds^2= -f(r)dt^2+\frac{dr^2}{g(r)}+ h(r)\left(d\theta^2+\sin^2\theta d\phi^2 \right),
    \label{eq:staticr}
\end{equation}
where
\begin{equation}
\begin{split}
    f(r)=g(r)&= 1- \frac{2M(r-d)-P^2-Q^2}{r^2-2dr-k^2}\\
    &= \left(1-\frac{2(d+M)}{r}+\frac{2P^2-k^2}{r^2}\right)\left(1-\frac{2d}{r}-\frac{k^2}{r^2}\right)^{-1}
    ,
    \label{eq:f func}
    \end{split}
\end{equation}
and
\begin{equation}
\begin{split}
    h(r)= R^2(r)&=r^2-2dr-k^2\\
    &=r^2\left(1-\frac{2d}{r}-\frac{k^2}{r^2}\right).
    \label{eq:h func}
    \end{split}
\end{equation}
In terms of the advanced Eddington-Finkelstein coordinates ($u,r,\theta,\phi$), where $du=dt- dr/f(r)$, Eq.~(\ref{eq:staticr}) is written as
\begin{equation}
    ds^2=-f(r)du^2-2du dr+ h(r)\left(d\theta^2+\sin^2\theta d\phi^2 \right).
    \label{eq:static_advance_null}
\end{equation}

The inverse metric components of the line element (\ref{eq:static_advance_null}) can be decomposed using the null tetrad $Z^{\mu}{}_{\alpha}=\left( l^{\mu}, n^{\mu},m^{\mu},\bar{m}^{\mu} \right)$ as 
\begin{equation}
    g^{\mu\nu}= -l^{\mu}n^{\nu}-l^{\nu}n^{\mu} +m^{\mu}\bar{m}^{\nu}+m^{\nu}\bar{m}^{\mu},
\end{equation}
where 
\begin{equation}
    l^{\mu}= \delta^{\mu}{}_{r},\quad~ n^{\mu}=\delta^{\mu}_{u}-\frac{f}{2}\delta^{\mu}_r, \quad~ m^{\mu}= \frac{1}{\sqrt{2 h}} \left(\delta^{\mu}_{\theta} + \frac{i}{\sin \theta }\delta^{\mu}_{\phi}\right).
\end{equation}
Using the complex transformation
\begin{equation}
    r\rightarrow r'= r+ i a \cos \theta, \quad~ u\rightarrow u'= u- i a \cos \theta,
\end{equation}
where $a$ is the rotation parameter, and replacing the terms $r^2$ by $\hat{\rho}^2= r'r'^{*}=r^2+a^2 \cos^2\theta$ and $\frac{2}{r}$ by $(\frac{1}{r'}+\frac{1}{r'^{*}})=\frac{2r}{\hat{\rho}^2}$, we get the new metric in the Eddington-Finkelstein coordinates \cite{Drake:1998gf,Shaikh:2019fpu}
\begin{equation}
\begin{split}
ds^2&= - F(r,\theta)du^2 -2du dr +2 a\sin^2\theta \left[F(r,\theta)- 1\right]du d\phi + 2 a \sin^2\theta dr d\phi \\
& \quad~ + H(r,\theta) d\theta^2 + \sin^2\theta \left[ H(r,\theta) + a^2 \sin^2\theta (2- F)\right]d\phi^2,
\label{eq:rotating advance null coord}
\end{split}
\end{equation}
where $F(r,\theta)$ and $H(r,\theta)$ are complexified forms of $f(r)$ and $h(r)$ respectively. In our case using Eqs.~(\ref{eq:f func}), (\ref{eq:h func}) we get
\begin{eqnarray}
f(r) \rightarrow F(r,\theta) &= & \left(1-\frac{2(d+M)r}{\hat{\rho}^2}+\frac{2P^2-k^2}{\hat{\rho}^2}\right)\left(1-\frac{2d r}{\hat{\rho}^2}-\frac{k^2}{\hat{\rho}^2}\right)^{-1}, \\
h(r) \rightarrow H(r,\theta) &=& \hat{\rho}^2\left(1- \frac{2 d r}{\hat{\rho}^2} -\frac{k^2}{\hat{\rho}^2}\right).
\end{eqnarray}
In Boyer-Lindquist coordinates, the new metric Eq.~(\ref{eq:rotating advance null coord}) for the rotating black hole finally takes the form \cite{Shaikh:2019fpu} 
\begin{equation}
\begin{split}
    ds^2 &= -F dt^2 -2 a (1- F)\sin^2\theta dt d\phi + \frac{H}{F H + a^2\sin^2\theta } dr^2  + H d\theta^2 \\
    & \quad~ + \sin^2 \theta \left[H + a^2 \sin^2 \theta (2- F)\right] d\phi^2.
    \end{split}
    \label{eq:rotating BL coord gen}
\end{equation}
After simplification using the explicit forms of $F(r,\theta)$ and $H(r,\theta)$ in our case, we arrive at the line element for a rotating dyonic black hole in Boyer-Lindquist coordinates 
\begin{equation}
\begin{split}
    \text{d}s^2= & -\left(1- \frac{2M(r-d)-P^2-Q^2}{\hat{\Sigma}}\right)\text{d}t^2-\frac{2a\sin^2\theta}{\hat{\Sigma}}(2M(r-d)-P^2-Q^2)\text{d}t\text{d}\phi \\
    & + \left(r^2-2dr-k^2+a^2+\frac{a^2\sin^2\theta}{\hat{\Sigma}}\left(2M(r-d)-P^2-Q^2\right)\right)\sin^2\theta \text{d}\phi^2+\frac{\hat{\Sigma}}{\hat{\Delta}}\text{d}r^2 +\hat{\Sigma}\text{d}\theta^2
    \end{split}
    \label{metric:DKSBH}
\end{equation}
where the functions $\hat{\Delta}(r)$ and $\hat{\Sigma}(r,\theta)$ are 
\begin{equation}
\hat{\Delta}=r^2-2dr -2M(r-d)-k^2+a^2+P^2+Q^2, 
\label{eq:hatdelta}
\end{equation}
and
\begin{equation}
    \hat{\Sigma}=r^2-2dr-k^2+a^2\cos^2\theta.
\end{equation}
This metric was already derived in \cite{PhysRevD.50.7394}, \cite{Wu:2020mby} using a different method. This is known as the dyonic Kerr-Sen black hole spacetime. Here $M$ and  $a$ are the mass and rotation parameters of the black hole. $Q$ and $P$ are the electric and magnetic charges, respectively. $d=(P^2-Q^2)/{2M}$ and $k=PQ/M$ are the dilaton charge and axion charge respectively. If the magnetic charge of the black hole vanishes, i.e. $P=0$ then it reduces to the Kerr-Sen black hole. For the special case, $P=Q$, the dilaton charge vanishes, i.e. $d=0$, but axion charge $k\neq 0$. This is a distinguishing feature of dyonic Kerr-Sen black holes. 

\begin{figure}[!htbp]
\centering
\subfigure[]{\includegraphics[width=3.0in,angle=360]{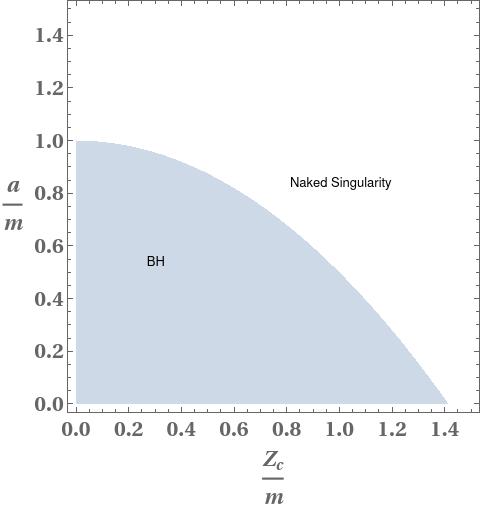}\label{subfig:f1}}
\subfigure[]{\includegraphics[width=3.0in,angle=360]{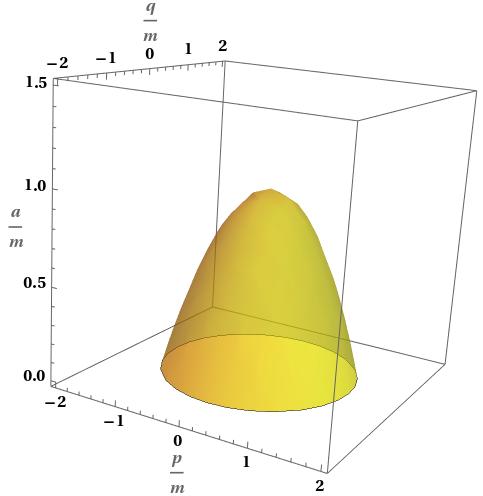}\label{subfig:f2}}
\caption{In (a) the shaded region in the $a/M$- $Z_c$ parameter space corresponds to black holes. In (b) the plot is extended into the full 3-D parameter space of $a/M$, $P/M$, and $Q/M$. Note that $Z_c^2=P^2+Q^2$.}
\label{fig:1}
\end{figure}

There is a curvature singularity at $r=0$ covered by the radius
\begin{equation}
    r_{\pm}= M+ \frac{P^2-Q^2}{2M}\pm \sqrt{\left(M-\frac{P^2+Q^2}{2M}\right)^2-a^2}.
    \label{eq:Rh}
\end{equation}
The corresponding event horizon and Cauchy horizon are given by $R_{+}=\sqrt{r_{+}^2-2d r_{+} -k^2}$ and $R_{-}=\sqrt{r_{-}^2-2d r_{-}-k^2}$ respectively. 
The horizon can exist only for
\begin{equation}
    \left(1-\frac{Z_c^2}{2M^2}\right)^2\geq \frac{a^2}{M^2},
\end{equation}
where $Z_c^2=P^2+Q^2$. Otherwise, the spacetime describes a naked singularity. This is illustrated in Fig.~\ref{fig:1}.

\newpage
\section{Black hole shadows}
In this section we study the shadow cast by the black holes (both rotating and non-rotating) on the observer's sky. 
We assume the black hole shadow in background light coming from distant
sources only.  In reality, other factors such as light coming from an accretion disk of the black hole or light propagation influenced by plasma and dust in the medium surrounding the black hole, should also be considered if  one is interested in the complete observational appearance of the shadow. However, the size and shape of the shadow which is of interest to us in this study, do not get significantly affected by these factors \cite{Perlick:2021aok}.

In order to separate the radial ($r$) and  angular ($\theta$) equation of motion for photons, we use the Hamilton-Jacobi method. For the rotating case we use Boyer-Lindquist coordinates. In non-rotating cases, Schwarzschild coordinates 
are the standard choice.

\subsection{Shadows of (rotating) dyonic Kerr-Sen black holes}
The Hamilton-Jacobi (HJ) equation for photon trajectories is given by
\begin{equation}
    H(x^{\mu},p_{\mu})+\frac{\partial S}{\partial \lambda}=0
\end{equation}
where $S(x^{\mu},\lambda)$ is the Jacobi action, $\lambda$ is the affine parameter, and $H(x^{\mu},p_{\mu})$ is the Hamiltonian corresponding to the Lagrangian null trajectories given by
$\mathcal{L}=\frac{1}{2}g_{\mu\nu}\dot{x}^{\mu}\dot{x}^{\nu}=0$, and ``dot" represents the derivative with respect to the affine parameter $\lambda$. The conjugate momentum is $p_\mu=\frac{\partial S}{\partial x^{\mu}}=\frac{\partial \mathcal{L}}{\partial \dot{x}^{\mu}}$. 

 We can write $S$, using separation of variables, as
\begin{equation}
S = -Et+L\phi + S^r(r)+S^{\theta}(\theta),
\end{equation}
where $E=-p_t$ and $L=p_{\phi}$ are the constants of motion. 
As $S$ does not depend explicitly on $\lambda$, the HJ equation becomes $H=\frac{1}{2}g^{\mu\nu}p_\mu p_\nu=0$.
Using $p_r=\frac{\partial S}{\partial r}=\frac{\D S^r}{\D r}$ and $p_{\theta}=\frac{\partial S}{\partial \theta}=\frac{\D S^{\theta}}{\D \theta}$ and the inverse metric components
\begin{eqnarray}
    g^{tt}&=&-\frac{\left(r^2-2dr-k^2+a^2\right)^2-\hat{\Delta}a^2\sin^2\theta }{\hat{\Sigma}\hat{\Delta}},\\
    g^{t\phi}&=&g^{\phi t}= - \frac{a}{\hat{\Sigma}\hat{\Delta}}\left(2M(r-d)-P^2-Q^2\right),\\
    g^{\phi\phi}&=& \frac{\hat{\Delta}-a^2\sin^2\theta}{\hat{\Sigma}\hat{\Delta}\sin^2\theta },\\
    g^{rr}&=& \frac{\hat{\Delta}}{\hat{\Sigma}}, \quad~ g^{\theta\theta}=\frac{1}{\hat{\Sigma}},
\end{eqnarray}
we expand the HJ equation and obtain the separated  angular and radial equations of motion for photons. The angular equation of motion is
\begin{equation}
\begin{split}
   \frac{\D S^{\theta}}{\D \theta} = & E\sqrt{\Theta(\theta)}\\
   =& E \sqrt{\chi -l^2\cot^2\theta +a^2 \cos^2\theta},\\
   \end{split}
\end{equation}
where $\chi=C/E^2$ ($C$ is the Carter constant) and $l= L/E$. The radial equation is
\begin{equation}
    \frac{\D S^{r}}{\D r} = E\sqrt{-V(r)},
\end{equation}
where the effective potential 
\begin{equation}
    V(r)= \frac{(l-a)^2+ \chi}{\hat{\Delta}}-\frac{(r^2-2d r -k^2 +a^2 -a l)^2}{\hat{\Delta}^2}.
\end{equation}
For unstable photon orbits $V(r_{ph})=V'(r_{ph})=0$. After simplification , we obtain $l(r_{ph})$ and $\chi(r_{ph})$ as
\begin{eqnarray}
    l(r_{ph})&=& \frac{1}{a}\left(r^2_{ph}+a^2-2 d r_{ph} - k^2 -4(r_{ph}-d)\frac{\hat \Delta (r_{ph})}{\hat \Delta^{\prime} (r_{ph})}\right), \label{eq:l(rph)} \\
    \chi(r_{ph}) &=& \frac{16(r_{ph}-d)^2 \hat{\Delta}(r_{ph})}{\hat{\Delta}^{\prime 2}(r_{ph})}-\frac{1}{a^2}\left(r^2_{ph}-2 d r_{ph} - k^2 -4(r_{ph}-d)\frac{\hat \Delta (r_{ph})}{\hat \Delta^{\prime} (r_{ph})}\right)^2, \nonumber\\
    \label{eq:chi(rph)}
\end{eqnarray}
where $\hat{\Delta}(r_{ph})$ and $\hat{\Delta}^{\prime}(r_{ph})$ can be obtained from Eq.~(\ref{eq:hatdelta}) . 

\begin{figure}[!htbp]
\centering
\includegraphics[width=10cm]{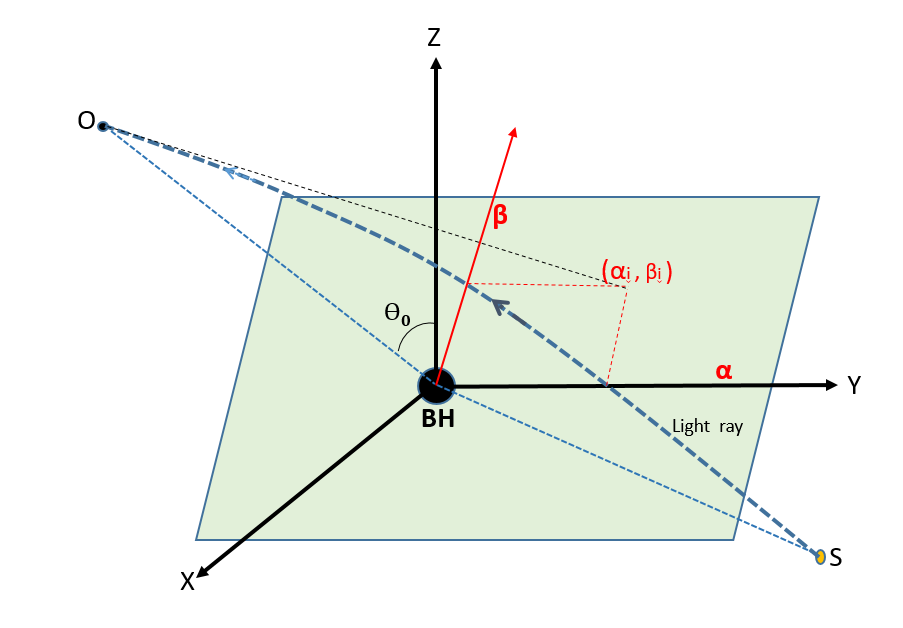}
\caption{Bardeen's ($\alpha,\beta$) coordinates for the observer's sky}
\label{fig:celestial_coord} 
\end{figure}

To observe the shadow on the sky, one needs to set up a suitable coordinate system. The four velocity of a timelike observer around a rotating black hole (such as Kerr spacetime) is given by \cite{Perlick:2021aok}
\begin{equation}
    u= \frac{(r^2+a^2)\partial_t + a\partial_{\phi}}{\sqrt{\Sigma \Delta}}\Big\vert_{r_O,\vartheta_O},
\end{equation}
where $r_O$ and $\vartheta_O$ are the coordinates of the observer with respect to a Kerr black hole expressed in Boyer-Lindquist coordinates. For the dyonic Kerr-Sen black hole, $\sqrt{\Sigma \Delta}$ in the expression is just replaced by $\sqrt{\hat{\Sigma}\hat{ \Delta}}$. We assume for the observer on Earth, $r_O>>M>a$. Then the observer four velocity takes the form $u\approx \partial_t$, i.e. a static observer. In general, the location and motion of the observer affects the size and shape of the shadow in the observer's sky. However, for the distant black holes M87* and SgrA* these effects are negligible, as discussed in \cite{Chang_2021}. For such static observers in an asymptotically flat spacetime, we use the celestial coordinates ($\alpha,\beta$) on the observer's sky. These coordinates 
were introduced by Bardeen \cite{Bardeen:1973tla,Vazquez:2003zm} (see Fig.~\ref{fig:celestial_coord}).

In Bardeen's coordinate system, the Boyer-Lindquist
coordinates describing the black hole spacetime coincide with the Cartesian XYZ
coordinates, at very large distances from the origin (the location of the black
hole). For such a distant observer, the sky is the $\alpha$-$\beta$ plane which is perpendicular to the line joining the observer to the black hole. For a light ray reaching the observer, the tangent drawn from the observer hits the sky at the point ($\alpha_i$, $\beta_i$). Thus, $\alpha$ and $\beta$ have the dimension of length (or mass, for $G=1,c=1$). Hence, to get the angles which define the shadow boundary, one has to use  $(\alpha/r_O,\beta/r_O)$, where $r_O$ is the distance to the black hole from the observer. Under these assumptions, the celestial coordinates for the observer's sky are obtained as \cite{Vazquez:2003zm}
\begin{equation}
    \alpha = -\frac{l}{\sin \theta_0}, \quad~ \beta = \pm \sqrt{\Theta(\theta_0)},
    \label{eq: celestial coord}
\end{equation}
where $\theta_0$ is the inclination angle of the observer with respect to the black hole's rotation axis Z. The  parametric plot $\alpha(r_{ph})$ versus $\beta(r_{ph})$ using the Eqs.~(\ref{eq:l(rph)}), (\ref{eq:chi(rph)}), and (\ref{eq: celestial coord}) gives the shadow profile.

However, for asymptotically non-flat spacetimes, one needs to use different coordinates. One such coordinate system  \cite{PhysRevD.89.124004,Grenzebach:2015oea} uses two astronomical angles, azimuthal angle ($\psi$) and co-latitude angle ($\vartheta$),  to locate a point on the celestial sphere with the observer at the origin.  The angle $\vartheta$ is the angle between the tangent to the light ray and the radial direction from the observer to the black hole. The angle $\psi$ is the azimuthal angle of the tangent ray in the equatorial plane orthogonal to the 
observer--black hole direction. Each light ray has a definite set of values ($\vartheta,\psi$). The shadow is defined by all $(\vartheta(r_{ph}),\psi(r_{ph}))$. The stereographic projection of all $(\vartheta(r_{ph}),\psi(r_{ph}))$ gives the shadow boundary. The above characterisation as well as its relation with the original Bardeen formalism is elegantly discussed in the well-known and very recent review \cite{Perlick:2021aok}.  

\begin{figure}[!htbp]
\centering
\subfigure[{\tiny $a/M=0.5,\, Q/M=0.85,\, P/M=0.25$}]{\includegraphics[width=2.1in,angle=360]{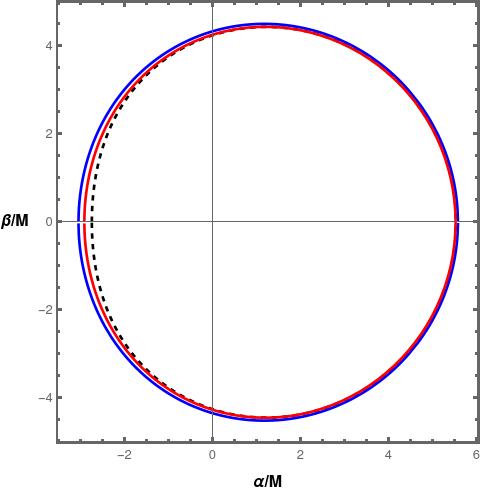}\label{subfig:f3}}
\subfigure[{\tiny $a/M=0.5,\, Q/M=0.85,\, P/M=0.35$}]{\includegraphics[width=2.1in,angle=360]{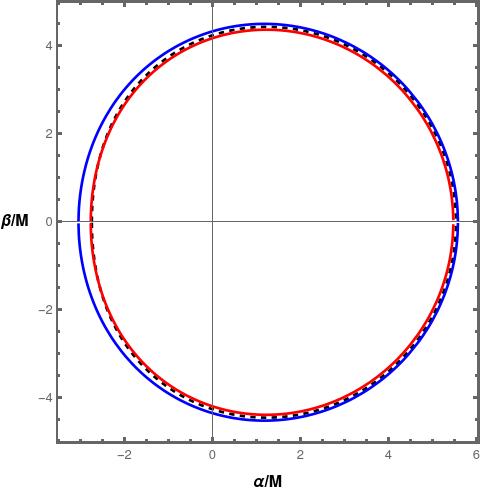}\label{subfig:f4}}
\subfigure[{\tiny $a/M=0.5,\, Q/M=0.85,\, P/M=0.45$}]{\includegraphics[width=2.1in,angle=360]{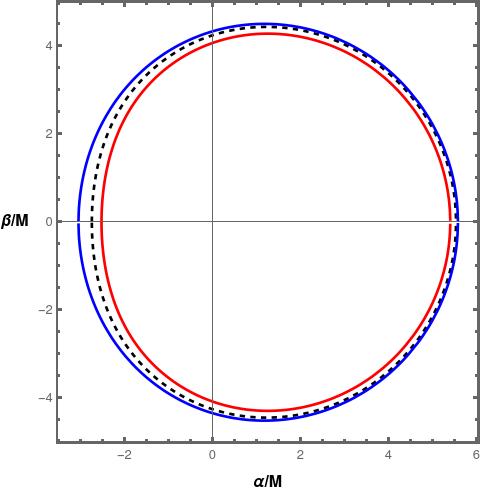}\label{subfig:f5}}
\subfigure[{\tiny $a/M=0.65,\, Q/M=0.75,\, P/M=0.20$}]{\includegraphics[width=2.1in,angle=360]{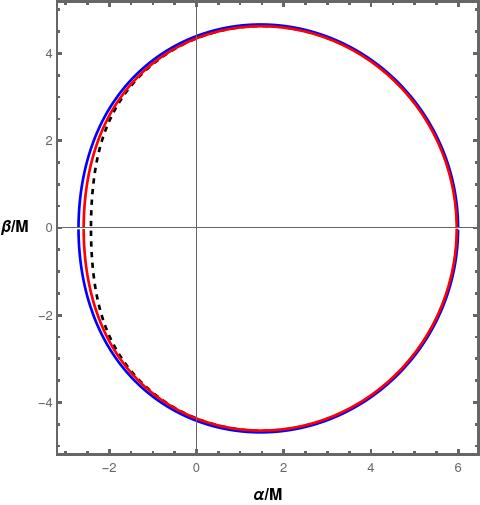}\label{subfig:f6}}
\subfigure[{\tiny $a/M=0.65,\, Q/M=0.75,\, P/M=0.25$}]{\includegraphics[width=2.1in,angle=360]{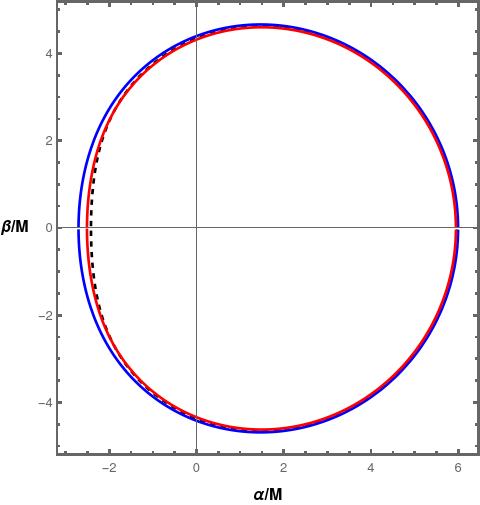}\label{subfig:f7}}
\subfigure[{\tiny $a/M=0.65,\, Q/M=0.75,\, P/M=0.35$}]{\includegraphics[width=2.1in,angle=360]{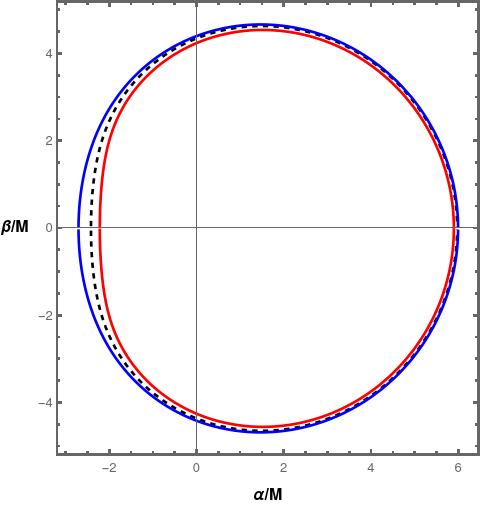}\label{subfig:f8}}
\caption{Shadow boundaries are plotted in the observer's sky, i.e. $\frac{\alpha}{M}$ vs $\frac{\beta}{M}$ space. The black dashed and the solid blue curves in each part denote the shadow boundaries of the Kerr-Newmann and the Kerr-Sen black holes, respectively. The solid red curve denotes the dyonic Kerr-Sen black holes. In the top panel, $a/M=0.5$ and $Q/M=0.85$ for all three parts ($(a),\,(b),\, (c)$) but $P/M$ increases for the dyonic Kerr-Sen black holes as we go from $(a)$ to $(c)$. In the bottom panel, the $a/M$ value is increased to $a/M=0.65$ for parts $(d)$ to $(f)$. The $Q/M$ value is also fixed at $Q/M=0.75$ but $P/M$ values are increased for dyonic Kerr-Sen black holes as in the top panel. The inclination angle $\theta_0= 90$ degrees for all of the parts.}
\label{fig:3}
\end{figure}

\begin{figure}[!htbp]
\centering
\subfigure[{\tiny $a/M=0.75,\, Q/M=0.65,\, P/M=0.27$}]{\includegraphics[width=2.1in,angle=360]{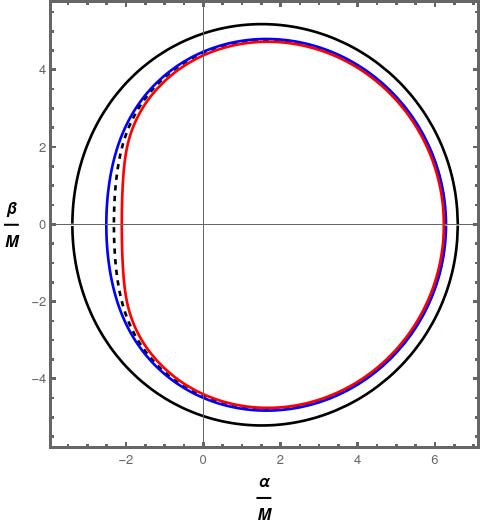}\label{subfig:f9}}
\subfigure[{\tiny $a/M=0.85,\, Q/M=0.52,\, P/M=0.165$}]{\includegraphics[width=2.1in,angle=360]{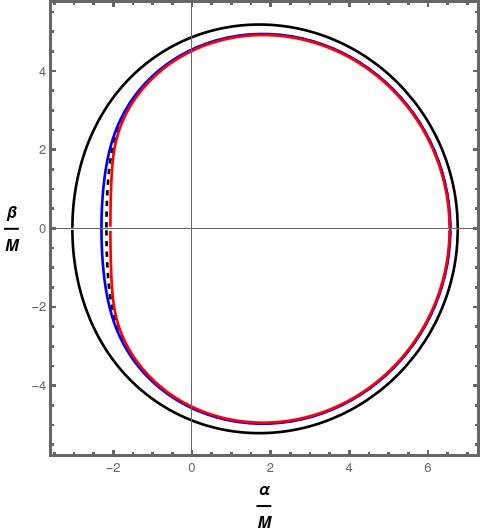}\label{subfig:f10}}
\subfigure[{\tiny $a/M=0.95,\, Q/M=0.30,\, P/M=0.09$}]{\includegraphics[width=2.1in,angle=360]{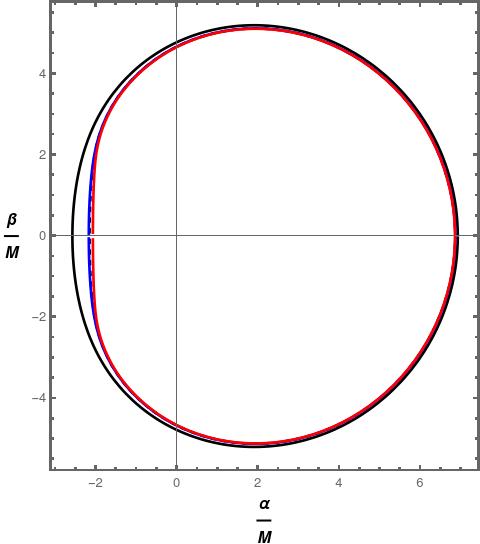}\label{subfig:f11}}
\caption{Shadow boundaries are plotted in the observer's sky, i.e. $\frac{\alpha}{M}$ vs $\frac{\beta}{M}$ space. The black dashed and the solid blue curves in each part denote the shadow boundaries of the Kerr-Newmann and the Kerr-Sen black holes, respectively. The solid red curve denotes 
values for the dyonic Kerr-Sen black holes. The outer black solid curve 
is for Kerr black holes with corresponding rotation parameter ($a/M$) value. The inclination angle $\theta_0= 90$ degrees for all of the parts.    }
\label{fig:4}
\end{figure}

In Fig.~\ref{fig:3}, the comparison between the shadow profiles for different values of $a/M$, $Q/M$, and $P/M$ are shown for the Kerr-Newmann, Kerr-Sen, and dyonic Kerr-Sen black holes. We note that as we increase the $P/M$ values the deviations from the Kerr-Sen and the Kerr-Newmann black holes are more prominent. However, as the $a/M$ value (rotation parameter) is increased the maximum deviation from the Kerr-Sen black holes is found to decrease. This is more prominent in Fig.~\ref{fig:4}, where all the shadow boundaries approach the outermost black solid curve for the Kerr black holes, as the rotation parameter value is increased. 

\subsection{Shadows of static black holes}
Using the line element [Eq.~(\ref{eq:static line element})] for static black holes, the equations of motion for the photon trajectories are
\begin{eqnarray}
     \frac{\D S^{\theta}}{\D \theta}= E \sqrt{\chi - l^2\cot^2\theta},\\
    \frac{\D S^R}{\D R }= E\sqrt{-V(R)},
\end{eqnarray}
where $\chi=C/E^2$, $l= L/E$, $C$ is Carter constant,  $L$ is angular momentum. The effective potential for the radial equation of motion 
\begin{equation}
    V(R)= \frac{\psi^2}{\Delta^2} \left[\frac{\chi + l^2}{R^2}-\frac{1}{\Delta^2}\right].
\end{equation}
For the photon sphere radius (corresponding to the unstable orbits), $V(R_{ph})=V'(R_{ph})=0$, which leads to the relation
\begin{equation}
    R_{ph}= \frac{\Delta(R_{ph}) }{\Delta'(R_{ph})}.
\end{equation}
Using the expression of $\Delta(R)$ [Eq.~(\ref{eq: Delta})], we obtain a quadratic equation 
\begin{equation}
\begin{split}
    x^2+ b x + c =& 0,\\
 \text{where}, \quad~   b=& \frac{\tilde{Z}_c^{4}}{4} + 4 \tilde{Z}_c^2 -9,\\ 
 \text{and}, \quad~ c = & \tilde{Z}_c^6- 2\tilde{Z}_c^4,
    \end{split}
\end{equation}
where $x=R_{ph}^2/M^2$ and $\tilde{Z}_c^2= (P^2+Q^2)/M^2$. Then we obtain the photon radius from the root of the quadratic equation, i.e. 
\begin{equation}
    x= \frac{R_{ph}^2}{M^2}= \frac{-b+ \sqrt{b^2- 4c}}{2}.
    \label{eq:x root}
\end{equation}
For the celestial coordinates  $\alpha$ and $\beta$ as defined earlier [Eq.~(\ref{eq: celestial coord})] with $a=0$, we obtain
\begin{equation}
    \alpha^2 + \beta^2 = \chi + l^2 = \frac{R^2_{ph}}{\Delta^2(R_{ph})}= \frac{1}{\Delta'^2(R_{ph})}.
\end{equation}
Therefore the shadow radius for static black holes is 
\begin{equation}
    \begin{split}
        R_{shadow}=& \frac{R_{ph}}{\Delta(R_{ph})} \\
        = & M \left[ \frac{x^2}{x-2\sqrt{x+ \frac{ \tilde{Z}_c^4}{4}} + \tilde{Z}_c^2 }\right]^{1/2},
    \end{split}
    \label{eq: shadow radius DKSBH}
\end{equation}
where $x$ is given by Eq.~(\ref{eq:x root}). For the critical value $\tilde{Z}_c^2=2$ the photon-sphere radius vanishes ($R_{ph}=0$) but the shadow radius does not vanish ($R_{shadow}=2M$). Thus, the shadow does not exist for naked singularities. For, $P=Q=0$, i.e. $\tilde{Z}_c=0$, $R_{shadow}=3\sqrt{3}M$, which is the case for the Schwarzschild black hole.

\section{Observational bound on rotating black holes}
We can test the possible existence of rotating dyonic Kerr-Sen black holes 
using the observations of black hole shadows of M87$^*$ and Sgr A$^*$. To do this, we may define two observational quantities which are-- $(i)$ deviation from circularity ($\Delta C$) \cite{Bambi:2019tjh} and $(ii)$ fractional deviation parameter ($\delta$) related to the average shadow diameter \cite{EHT2022_6,Afrin:2022ztr,Shaikh:2022ivr}. These two quantities are described as follows. 

We note that the shadow profile is symmetric about $\beta=0$, i.e. the $\alpha$ axis. The geometric centre of the shadow image on the $\alpha$- axis is obtained by taking its mean. Therefore, the centre of the shadow profile is
\begin{equation}
    \alpha_c = \frac{\int \alpha\, \D A}{\int\, \D A}, \quad~ \beta_c =0,
\end{equation}
where $\D A = 2\beta d\alpha$ is the area element on the shadow image. 
From the geometric centre of the shadow image, the radial distance $\ell(\phi)$ to any point on the shadow boundary, making an angle $\phi$ with respect to the $\alpha$- axis, can be expressed as
\begin{equation}
   \ell(\phi)= \sqrt{\left(\alpha(\phi)-\alpha_c\right)^2+\beta(\phi)^2}. 
\end{equation}
Then the average shadow radius can be defined as the root mean squared radius, i.e. 
\begin{equation}
    R_{avg}^2= \frac{1}{2\pi} \int^{2\pi}_0 \D \phi \ell^2(\phi).
\end{equation}
Finally, the deviation from circularity is defined as \cite{Bambi:2019tjh}
\begin{equation}
    \Delta C = \frac{1}{R_{avg}}\sqrt{\frac{1}{2\pi}\int^{2\pi}_0 \D \phi \left(\ell(\phi)-R_{avg}\right)^2}.
\end{equation}

For the computation, it is more convenient to use $r_{ph}$ as the parameter instead of $\phi$. Then, we can express $R_{avg}$ and $\Delta C$ as 
\begin{eqnarray}
    R^2_{avg}&=& \frac{1}{\pi} \int^{r_{ph-}}_{r_{ph+}} \left(\beta'(\alpha -\alpha_c)-\beta \alpha'\right)\D r_{ph},\\
    \Delta C&=& \frac{1}{R_{avg}} \sqrt{\frac{1}{\pi} \int^{r_{ph-}}_{r_{ph+}} \left(\beta'(\alpha -\alpha_c)-\beta \alpha'\right)\left(1-\frac{R_{avg}}{\ell}\right)^2\D r_{ph} } ,
\end{eqnarray}
where $\beta'=\frac{\D \beta}{\D r_{ph}}$ and $\alpha'=\frac{\D \beta}{\D r_{ph}}$. Here $r_{ph+}$ and $r_{ph-}$ are obtained from the roots of $\beta(r_{ph})=0$, i.e. the values of $r_{ph}$ for which the shadow boundary cuts the $\alpha$-axis. In other words, $\phi(r_{ph+})=0$ and $\phi(r_{ph-})=\pi$. The geometric centre of the shadow ($\alpha_c$, $\beta_c$) can also be expressed in terms of the parameter $r_{ph}$ as
\begin{equation}
    \alpha_c= \frac{\int^{r_{ph-}}_{r_{ph+}}\alpha \beta \alpha' \D r_{ph}}{\int^{r_{ph-}}_{r_{ph+}} \beta \alpha' \D r_{ph}}, \quad~ \beta_c= 0.
\end{equation}
Note that $\alpha(r_{ph})$, $\beta(r_{ph})$ are obtained from Eqs.~(\ref{eq:l(rph)}), (\ref{eq:chi(rph)}), and (\ref{eq: celestial coord}).

 Six parameters are used to
describe a dyonic Kerr-Sen black hole solution. 
These are mass ($M$), rotation ($a$), electric charge ($Q$), magnetic charge ($P$), dilaton charge ($d$), and  axion charge ($k$). However, the dilaton and axion charges depend on the parameters $P$ and $Q$, as $d= \frac{P^2-Q^2}{2M}$ and $k=\frac{PQ}{M}$. The black hole mass $M$ is determined by other observations. The geometry of the spacetime and, consequently, the shadow profile possess a symmetry so that instead of treating $P$ and $Q$ as independent parameters we define a new parameter $Z_c^2=P^2+Q^2$. All possible values of $P$ and $Q$ satisfying a given fixed value of $Z_c$ give the same geometry and the shadow profile. This is due to the duality of the dyonic charges. Thus effectively, we have two free parameters: the rotation $a$ and $Z_c$. We constrain this parameter space ($a/M$ vs. $Z_c/M$) from observations of the shadows of M87* and SgrA*.

\begin{figure}[!htbp]
\centering
\subfigure[]{\includegraphics[width=3.2in,angle=360]{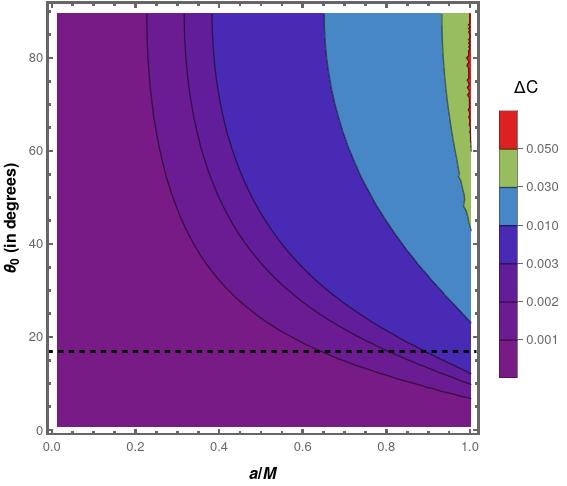}\label{subfig:f12}}
\subfigure[]{\includegraphics[width=3.2in,angle=360]{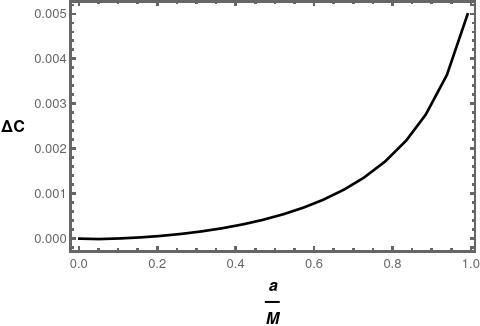}\label{subfig:f13}}
\caption{ $(a)$ The contour plots for different values of $\Delta C$ are shown over $a/M$ vs. $\theta_0$ parameter space, for Kerr black holes. The black dashed line corresponds to $\theta_0=17^{o}$. $(b)$ $\Delta C$ is plotted as a function of $a/M$ for the inclination angle $\theta_0=17^{o}$.   }
\label{fig:5}
\end{figure}

\begin{figure}[!htbp]
\centering
\subfigure[]{\includegraphics[width=3.2in,angle=360]{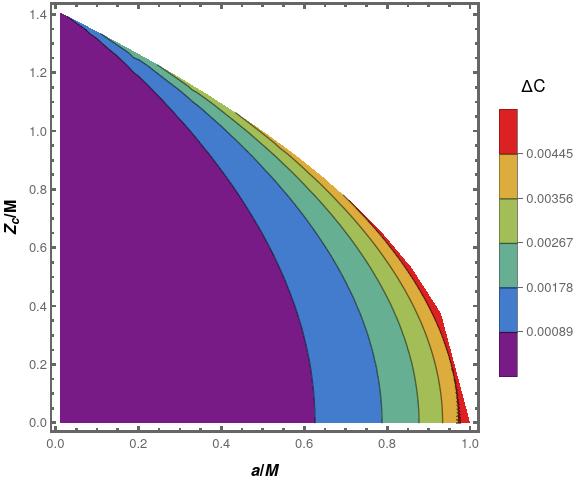}\label{subfig:f14}}
\subfigure[]{\includegraphics[width=3.2in,angle=360]{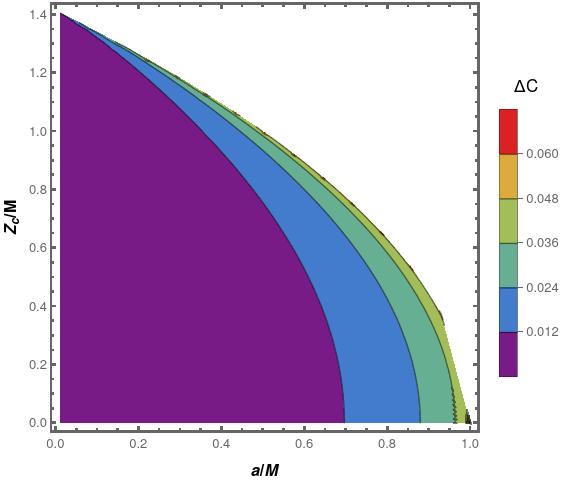}\label{subfig:f15}}
\caption{ The contour plots for different values of $\Delta C$ are shown over $a/M$ vs. $Z_c/M$ parameter space for dyonic Kerr-Sen black holes. For plot $(a)$ the inclination angle is $\theta_0=17^{o}$ and for plot $(b)$ $\theta_0= 90^{o}$. In the plots, $Z_c=\sqrt{P^2+Q^2}$. The white excluded region of the parameter space is for naked singularities. }
\label{fig:6}
\end{figure}

From the observation of the shadow of M87$^*$, the EHT collaboration has given a bound $\Delta C \lesssim 0.1$ for an inclination angle $\theta_0= 17^{o}$ \cite{Akiyama_2019L1,Akiyama_2019L5,Akiyama_2019L6}. However, the bound on $\Delta C$ from SgrA$^*$ is not available. From Fig.~\ref{subfig:f12}, we note that for Kerr black holes the maximum value $\Delta C \lesssim 0.07 $ for all inclination angles ($\theta_0$). Considering the orientation of the observed relativistic jets from the M87$^{*}$, the inclination angle is estimated to be $\theta_0 =17^{o}$ \cite{Walker_2018}. In Fig.~\ref{subfig:f13}, the variation of $\Delta C$ is shown as the function of $a/M$. Note that the maximum value of $\Delta C \lesssim 0.005$ for the inclination angle $\theta_0 =17^{o}$. For the same inclination angle, we scanned the parameter space $a/M - Z_c/M$ for the dyonic Kerr-Sen black holes. From Fig.~\ref{subfig:f14} we note that the for $\Delta C \lesssim 0.00534$. As $\Delta C$ increases with the inclination angle $\theta_0$, we have also scanned the parameter space for the inclination angle $\theta_0= 90^{o}$ in Fig.~\ref{subfig:f15}. The maximum possible deviation is $\Delta C \lesssim  0.072$. Therefore, we conclude that all black hole parameters are allowed and the present observational bound on the deviation from circularity cannot constrain the parameter space of the dyonic Kerr-Sen black holes.  

The recent EHT papers on SgrA$^*$ observations have used the fractional deviation parameter $\delta$ to constrain the spacetime geometries different from the Schwarzschild or Kerr black holes. The definition of $\delta$ is as follows
\begin{equation}
    \delta = \frac{d_{sh}}{d_{sh,Sch}}-1 =\frac{R_{avg}}{3\sqrt{3}M} -1 ,
\end{equation}
 where the average diameter of the shadow , $d_{sh}=2R_{avg}$. Using the observations of the shadow of SgrA$^*$ and two separate sets of prior values of mass and distance of SgrA$^*$ from the VLTI and Keck observations, the EHT  collaboration provided the bound on $\delta$ \cite{EHT2022_1,EHT2022_6} as
 \begin{equation}
 \delta= \begin{cases} 
      -0.08^{+0.09}_{-0.09}& \text{(VLTI)} \\
      -0.04^{+0.09}_{-0.10} & \text{(Keck)}  \\
   \end{cases}
\end{equation}
Therefore, we get the common range of $\delta$,  $-0.14<\delta < 0.01$, which is in the observational limits of both VLTI and Keck data.

\begin{figure}[!htbp]
\centering
\includegraphics[width=3.0in,angle=360]{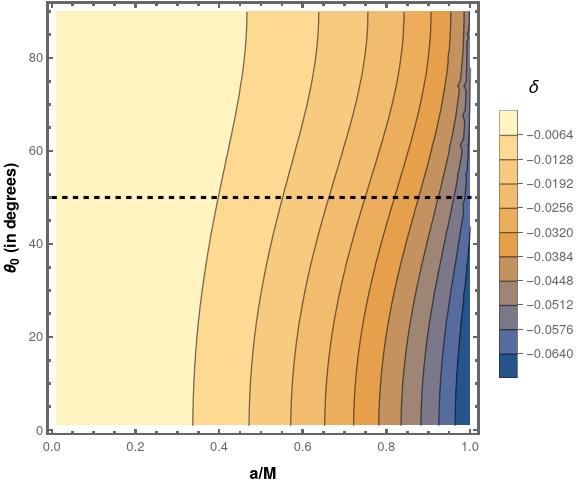}
\caption{The contour plots for different values of $\delta$ are shown over $a/M$ vs. $\theta_0$ parameter space, for Kerr black holes. The black dashed line corresponds to $\theta_0=50^{o}$.}
\label{fig:7}
\end{figure}

\begin{figure}[!htbp]
\centering
\subfigure[$\theta_0=50^{o}$]{\includegraphics[width=3.2in,angle=360]{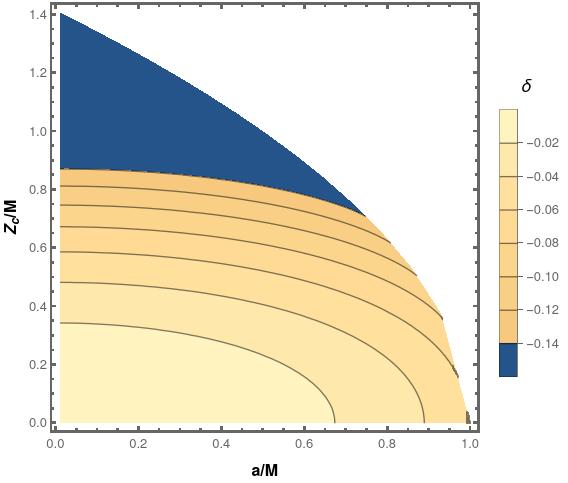}\label{subfig:f16}}
\subfigure[$\theta_0=50^{o}$, $a=0$]{\includegraphics[width=3.2in,angle=360]{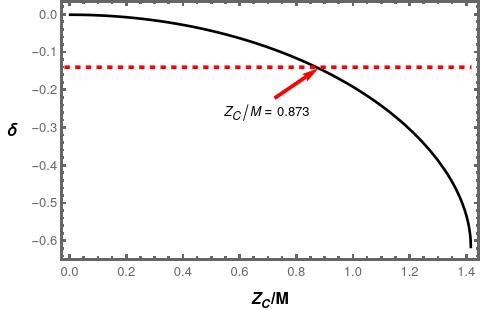}\label{subfig:f17}}
\caption{ $(a)$ The contour plots for different values of $\delta$ is shown over $a/M$ vs. $Z_c/M$ parameter space for dyonic Kerr-Sen black holes. The inclination angle is $\theta_0=50^{o}$. The blue shaded region is observationally disfavored as there $\delta< -0.14$. $(b)$ The black solid line is the plot of $\delta $ as the function of $Z_c/M$ for static black holes, i.e. $a=0$. The red dashed line corresponds to $\delta= -0.14$ (the observational limit). It intersects the black solid curve at $Z_C/M = 0.873$, meaning $Z_C/M\lesssim 0.873$ to satisfy the observational constraint.  }
\label{fig:8}
\end{figure}

In Fig.~\ref{fig:7}, we scan the parameter space $a/M - \theta_0$ with the contours labeled by different values of $\delta$ for the Kerr black holes. It is noted that $-0.0704 \lesssim \delta <0 $ for all parameter values. Thus, the Kerr black hole parameters are unconstrained from the observational bound on $\delta$ from SgrA$^{*}$. Further, we note that the variation in $\delta$ is less sensitive to the variation of $\theta_0$.  Moreover, in the observation of SgrA$^{*}$, inclination angle greater than $50^{o}$ is disfavored.

Therefore, we choose $\theta_0 =50^{o}$ in Fig.~\ref{subfig:f16}, where we scan the parameter space $a/M- Z_c/M$ with contours of different $\delta$ values for the dyonic Kerr-Sen black holes. In the blue shaded region of the plot $\delta< -0.14$ which means that the corresponding 
parameter values ($a/M, Z_c/M$) are not allowed according to the observations of the SgrA$^{*}$. Further, we note that for a $Z_c/M$
value greater than the critical limit $0.873$, no $a/M$ 
value can satisfy the observational constraint. This critical value of $Z_c/M$ is independent of the inclination angle $\theta_0$ as the critical limit corresponds to the $a=0$, i.e. the static black hole. In Fig.~\ref{subfig:f17}, we show the variation of $\delta$ as a function of $Z_c/M$ for static black holes, using the analytical expression for the shadow radius given in Eq.~(\ref{eq: shadow radius DKSBH}). There, we explicitly show the critical limit of $Z_c/M$. We conclude that $Z_c/M\lesssim 0.873$ and for any of allowed $Z_c/M$ values, the allowed range of $a/M$ must be within the allowed region of space shown in Fig.~\ref{subfig:f16}. In other words, a value $Z_c/M$ $> $ 0.873 is not allowed for any rotation parameter $a$.

\section{Comparison with other observations}
Magnetic monopoles are a natural outcome of grand-unified theories (GUTs) \cite{HOOFT1974276} and were among the original motivations for cosmic inflation \cite{PhysRevD.23.347,PhysRevLett.43.1365}. They have been searched for in various experiments \cite{MACRO:2002jdv,Orito:1990ny,Balestra:2008ps}. Primordial black holes with magnetic charges, which have reached the extremality condition in the course of cosmic evolution and do not Hawking radiate anymore, can be a possible dark matter candidate \cite{Diamond:2021scl}.  They are termed as the extremal Magnetic Black Holes (EMBHs). Various astrophysical limits on such EMBHs were discussed in \cite{Diamond:2021scl}. The authors in
\cite{Diamond:2021scl}  put constraints on such black holes with mass range having the upper limit $M<10^{33}\, \text{gm}$. Therefore, these black holes cannot be used for modeling supermassive black holes such as M87$^*$ or SgrA$^*$. Hence, bounds obtained from observations of shadow images, as discussed in this article
are not directly applicable to such EMBHs. However, we can 
develop some qualitative idea on what type of physical processes can indeed give bounds on 
any magnetic charge present in supermassive black holes 
such as the one at the center of our galaxy. In the following, we discuss two such possibilities.

$(a)$ From the observed temperature of clouds of Warm Ionized Medium (WIM) in the Milky Way \cite{Lehner:2004ty}, we can put some upper bound on the magnetic charge of SgrA*. Following the calculations of \cite{Diamond:2021scl}, we get a rough estimate of a upper bound on any
magnetic charge that may be present in SgrA$^*$.
While the magnetized black hole passes through the surrounding plasma (WIM), the black hole deposits energy into the plasma. The rate of energy deposition is given by \cite{Diamond:2021scl}
\begin{equation}
    \frac{dE}{dt}= 9\times 10^6 \left(\frac{P}{10^9\, \text{gm}}\right) \, \text{erg/sec},
    \label{heating}
\end{equation}
where $P$ is the magnetic charge (in the mass unit) of the black hole. Note that, in \cite{Diamond:2021scl} natural units $c=1, \hslash=1$ are used. In these units, the magnetic charge $Q_m$ is unit less and given by $Q_m=P/m_p$ where the Planck mass $m_p=G^{-1/2}=1.22\times 10^{19}\, \text{GeV}$.  This process  heats up the surrounding gas clouds.
On the other hand, the observed cooling rate per unit volume for the WIM gas clouds in the Milky Way is
\begin{equation}
    \frac{dE}{dt}/\text{cm}^3 \sim n_H \times 10^{-25.65^{+0.11}_{-0.15}} \, \text{erg/sec},
\end{equation}
where $n_H\sim 0.25 \, \text{cm}^{-3}$ is the typical number density
of hydrogen atoms in the gas cloud. The low density WIM fills more than 20\%  of the  volume within a 2 Kpc thick layer
around the galactic midplane. These results have been derived from a relatively small number of sight lines that sample a region 
of radius 3-4 Kpc about the Sun \cite{reynolds1992warm}. 
Therefore, around SgrA*, consider a region of disk of radius 8 Kpc (i.e. the distance from the Sun to the SgrA*) and thickness 2 Kpc. We assume that WIM clouds fill 20\%  of this region of space. So, the effective volume of WIM, $V\sim 80.42 \, \text{Kpc}^3$. Therefore, the net cooling rate of WIM surrounding SgrA* up to the distance to the Sun is roughly 
$\left(\frac{dE}{dt}/\text{cm}^3\right)\times V \sim 1.21\times 10^{40} \text{erg/sec}$. (Here, note that the cooling rate per unit volume may be different as we go far away beyond 3-4 Kpc from the Sun and near  SgrA*. But for a rough estimate we can assume the cooling rate to be the same throughout the region of space we considered.) 
Now the heating rate given by Eq.~(\ref{heating}) must be less than the net cooling rate, i.e.  
\begin{equation}
    9\times 10^6 \left(\frac{P}{10^9 \text{gm}}\right)^2 \leq 1.21\times 10^{40}.
\end{equation}
Thus the bound on the magnetic charge $P\leq 3.66\times 10^{25}\,\text{gm}$ . SgrA* has mass $M=4.154\times 10^6 M_{\odot}$. Thus, $P/M\leq 4.4 \times 10^{-15}$.

$(b)$ Another astrophysical constraint comes from the Parker bound \cite{Parker} which is based upon the survival of today's galactic magnetic field, as the field energy is drained out by the magnetic monopoles while moving through the field. This puts an upper limit on the flux of magnetic monopoles. 
Monopoles moving through a magnetic field extract energy from the field at the rate $\Vec{j_M}.\Vec{B}$ causing dissipation of the field energy
in the characteristic time 
\begin{equation}
    \tau \simeq \frac{1}{8\pi} \frac{B^2}{\Vec{j_M}.\Vec{B}},
\end{equation}
where $\Vec{j_M}$ is the magnetic current density and $B\sim 3\times 10^{-6}$ Gauss is the galactic magnetic field. $\tau > 10^8$ years if the field can be regenerated in a time as short as $10^8$ years.
The ``Extended Parker bound" obtained  by requiring survival
and growth of a small galactic seed field after the collapse of the protogalaxy is \cite{PhysRevD.62.025002} 
\begin{equation}
    \mathcal{F}\leq 10^{-21}\left(\frac{m}{10^{17}\, GeV}\right) cm^{-2}sec^{-1}Sr^{-1},
\end{equation}
where $\mathcal{F}$ is the flux of magnetic monopoles around the solar system and $m\sim 10^{15}-10^{19}$ GeV. 
The magnetic current density $\vec{j}_M=\rho_M \Vec{v}$, where $\rho_M$ is the magnetic charge density and $\Vec{v}$ is velocity of the charges. The  flux of magnetic monopoles is  $\mathcal{F}=\frac{\vert \vec{j_M} \vert}{4\pi}$. The relative velocity of the SgrA* with respect to the Solar System is $v_{\odot}\sim 220$ km/sec and $\rho_M = \frac{P}{m_p}\delta^3(\Vec{r})$ for the magnetic monopole charge of SgrA*. Thus the effective flux of magnetic charge of SgrA* around the solar system is
\begin{equation}
    \mathcal{F}= \frac{3}{16\pi^2}\frac{P}{m_p}\frac{v_{\odot}}{R^3}\leq 10^{-23} cm^{-2} s^{-1}Sr^{-1},
\end{equation}
where $R=8Kpc$ (distance to the SgrA* from the Solar System).
Therefore, $P\leq 3.59\, M_{\odot}$ and $P/M\leq 10^{-6}$.


Clearly, the above astrophysical constraints 
seem to be far more stringent, by several orders of magnitude
than the constraint obtained from shadow observations (i.e.
the result $P\leq 0.873 M$).

\section{Discussion and conclusions}
In this paper we have revisited black holes with dyonic charges in Einstein-Maxwell--dilaton-axion (EMDA) theory in the context of the observations on shadows of the supermassive black holes M87$^*$ and SgrA$^*$. 

First we outlined the derivation of the static black hole solution by direct integration of the field equations of EMDA theory. Further, the rotating version 
(the dyonic Kerr-Sen black hole) was obtained from the static solution by applying the Newman-Janis algorithm. Thereafter,  we analyzed the structure of the static black holes in detail. The differences with the standard Reissner-Nordstr\"om (RN) black holes are parametrized by the quantity $Z_c^2=P^2+Q^2$. These static black holes have a single horizon (the event horizon) unlike non-extremal RN black holes. For either $P=0$ or $Q=0$ the axion field vanishes. However the dilaton field vanishes only when both $P=0$ and $Q=0$, which is the case for the Schwarzschild black hole. Thus there is no solution with only axion and Maxwell's electromagnetic field but without dilaton field. The reason lies in the structure of the EMDA action where the axion field is linearly coupled with $F_{\mu\nu}\tilde{F}^{\mu\nu}$. The static dyonic black holes satisfy all energy conditions outside the event horizon. There is also no stable photon orbit outside the event horizon of these static dyonic black holes.

For the rotating case ($a\neq 0$) the axion field is nonzero even if the axion charge is zero ($k= PQ/M=0$) which is the case for the Kerr-Sen black hole ($P=0$, $Q\neq0$, and $a\neq 0$). In the static limit of the Kerr-Sen black holes (i.e. setting $a=0$), the axion field vanishes. This is an interesting difference between the Kerr-Sen black holes and the dyonic Kerr-Sen black holes, in general.

We study the shadow profiles for both rotating and static black holes. We have obtained the exact expression for the shadow radius for static black holes.
The parameter $Z_c < \sqrt{2}M$ and the shadow radius $R_{shadow}$ 
obeys $2M <R_{shadow}< 3\sqrt{3}M$. The effect of magnetic charge/ monopole on the shadow profile of rotating black holes has been investigated graphically by comparing the dyonic Kerr-Sen black holes with the Kerr-Newman and  the Kerr-Sen black holes (Fig.~\ref{fig:3}). The deviation increases with the increase in magnetic charge. However, the deviation is not so prominent for higher rotation parameter (Fig.~\ref{fig:4}).

Finally, we test whether the known supermassive black holes at 
galactic centres could be modeled as such dyonic Kerr-Sen black holes. 
In other words, we look for metric parameter values for which the
shadow features match with those in the
 observed shadow images of M87$^*$ and SgrA$^*$. We have used two observational quantities related to black hole shadows -- $(i)$ the deviation from circularity of the observed shadow boundary ($\Delta C$) and $(ii)$ the fractional deviation parameter ($\delta$) representing the deviation of the observed average shadow diameter from that for
 a Schwarzschild black hole. The observational bound on $\Delta C$ for M87$^*$ is satisfied for all parameter values for the dyonic black holes. Thus, it cannot be constrained. For SgrA$^*$, the EHT collaboration provided a bound on $\delta$ which gives a constraint on the black hole parameter $Z_c=\sqrt{P^2+Q^2}\lesssim 0.873 M$. Thus we get an upper bound on the magnetic monopole charge (if any) for SgrA$^*$ as $P\lesssim 0.873 M$ where $M=4.154\times 10^6 M_{\odot}$ for SgrA$^*$. In natural units, a magnetic charge $Q_m= P/m_{P}$ where $m_P=G^{-1/2}=1.22\times 10^{19}$ $GeV$ is the Planck mass. In these units, the obtained bound on the magnetic charge of SgrA$^*$ is $Q_m\leq 3.33\times 10^{44}$.

In the literature any other observational constraint on the possibility of magnetic monopole charge of SgrA* does not exist. There are some astrophysical constraints on the Extremal Magnetic Black Holes (EMBHs), which are basically primordial black holes and dark matter candidates. But these constraints are not useful for SgrA* and M87* as the mass of the supermassive black holes are extremely high as compared to those EMBH mass ranges. However, in Section VI, we have estimated roughly the astrophysical constraints  on the magnetic monopole charge of  SgrA*, following arguments similar to those discussed in \cite{Diamond:2021scl}. From the observed cooling rate of the WIM clouds in the Milky Way, we can get a rough estimate $P/M\leq 4.4\times 10^{-15}$ and from the Parker bound on the flux of the magnetic charges, we get an estimate $P/M\leq 10^{-6}$. Both these estimates are more stringent by several orders of magnitude than those we obtained from the observed shadow of SgrA*. However, for M87* situated at a distance of 16.8 Mpc, similar astrophysical observations
do not exist and hence similar constraints (as found for
SgrA*) cannot be estimated.


Thus, at present, it is difficult to reach any definite conclusion about the viable presence/absence of magnetic monopole charges in the supermassive black holes such as M87* and SgrA*. Future observations of shadow images at greater experimental sensitivities and more concrete analysis of other astrophysical constraints may 
provide bounds which are closer in value to each other and hence
more conclusive in nature. However, the theoretical analysis we have carried out is indeed new and may be 
of use once more imaging observations on shadows of black holes are 
carried out and presented in future.

\section*{Acknowledgements}
The research of SJ is partially supported by the SERB, DST, Govt. of India, through a TARE fellowship
grant no. TAR/2021/000354, hosted by the department of Physics, Indian Institute of Technology Kharagpur.

\bibliographystyle{apsrev4-1}
\bibliography{reference.bib}
\end{document}